# Invisible to humans, visible to machines: a preregistered audit of Unicode fidelity across four biomedical bibliographic APIs

**Author:** Przemysław Czuma[a]

[a] Polish Association for Artificial Intelligence in Medicine, ul. Grabowa 4a/8, 43-178 Ornontowice, Poland

**Corresponding author:** Przemysław Czuma, Polish Association for Artificial Intelligence in Medicine, ul. Grabowa 4a/8, 43-178 Ornontowice, Poland. Email: przemek.czuma@gmail.com. ORCID: 0009-0009-8235-2053.

---

## Abstract

**Objective:** To develop a reusable, preregistered method for auditing the character-level fidelity of bibliographic application programming interfaces (APIs), and to apply it to four public APIs used to build biomedical text corpora.

**Methods:** Preregistered audit (OSF osf.io/zjevr). From the 2024 PubMed Central (PMC) Open Access subset, a simple random sample of 4,000 English-language research articles was drawn. Abstracts returned by the PubMed E-utilities, Crossref, OpenAlex and Semantic Scholar were compared with the PMC Journal Article Tag Suite (JATS) XML abstract for 51 registered code points, 34 of them in four primary classes: typographic punctuation, mathematical/scientific symbols, Greek letters and special whitespace. A cell (API × class) was classified as systematic loss when the upper 95% Wilson bound of conditional preservation was below 5%, and as faithful when the lower bound exceeded 90%. The mechanism of each loss was examined in the aligned text and the raw API responses.

**Results:** Two of 16 cells met the loss criterion. PubMed preserved typographic punctuation in 0.6% of eligible abstracts (12/2,113; 95% CI 0.3–1.0%), replacing it with ASCII equivalents. OpenAlex preserved special whitespace in 0.0% (0/1,085; 95% CI 0.0–0.4%), which its inverted-index format cannot store. Mathematical symbols and Greek letters were preserved by all four APIs (95.2–99.7%). Crossref returned an abstract for 75.4% of papers (95% CI 74.1–76.7%) and for none published by Elsevier (0/360) or the American Chemical Society (0/105).

**Conclusion:** The same published abstract reaches text-mining pipelines with different characters depending on the API that serves it. Studies that analyze characters, including character-level markers of writing assisted by large language models, should take text from a source shown to preserve those characters, or normalize all sources identically, and report the API and retrieval date.

---

# 1. Introduction

## 1.1 The problem

Biomedical text mining, the construction of training corpora for biomedical language models and scientometric analyses usually take abstracts from bibliographic application programming interfaces (APIs) rather than from publishers' full text. They assume that the API returns the published abstract. Between the publisher's deposit and the consuming tool, the text passes through indexing and serialization steps, each of which can change single characters without changing what a human reader sees. Tokenizers, string matching, hash-based deduplication and character counts, however, operate on code points, not on rendered glyphs: replacing the em dash U+2014 with the hyphen-minus U+002D, or the no-break space U+00A0 with an ordinary space, changes the bytes and therefore the token sequence and any statistic computed on characters or tokens.

## 1.2 The present study

This preregistered audit compares the abstracts returned by four APIs (PubMed E-utilities, Crossref, OpenAlex, Semantic Scholar) with the PubMed Central (PMC) Journal Article Tag Suite (JATS) XML abstract of the same paper, in a simple random sample of 4,000 English-language research articles from the 2024 PMC Open Access subset. For each paper and character class, it records whether at least one character of the class present in the JATS abstract is present in the abstract returned by each API. Two co-primary estimands separate fidelity from availability: conditional preservation, among papers for which the API returns an abstract, and end-to-end preserved availability, among all papers whose JATS abstract contains the class. For each cell classified as a systematic loss, the mechanism is audited. The Crossref coverage gap is reported separately, because it results from publisher deposit rather than from character handling. The audit design itself is the methodological contribution of the paper; it is outlined in §3.1, and its generalizability and range of applicability are discussed in §5.7.

**Statement of significance**

| | |
|---|---|
| **Problem or Issue** | Text-mining and corpus pipelines assume that bibliographic APIs return the published |
| | abstract character for character. |
| **What is Already Known** | Some publishers deposit no abstracts to Crossref, and OpenAlex serves abstracts only as an inverted index. To the author's knowledge, character-level fidelity has not been measured. |
| **What this Paper Adds** | A reusable, preregistered method for auditing character-level fidelity against a common reference, and its first application to four APIs: PubMed replaces typographic punctuation with ASCII, OpenAlex removes special whitespace, and Greek letters and mathematical symbols survive everywhere. |
| **Who would benefit from the new knowledge in this paper** | Builders of biomedical text corpora and researchers who count characters in abstracts. |

## 2. Related work

Studies of AI-assisted writing now measure features of abstract text at scale. Kobak et al. identified "excess vocabulary" in PubMed abstracts, words whose frequency rose after the release of ChatGPT [1]. Keck reported a rise in em-dash frequency in ecology abstracts retrieved from OpenAlex [2]. Liang et al. estimated the share of text modified by large language models (LLMs) in about 1.1 million papers on arXiv, bioRxiv and in Nature portfolio journals [3]. Each of these studies analyzes the text as its chosen source returns it. For whole words this is usually safe; for features carried by single, less common Unicode characters it has not been tested.

Comparisons of bibliographic databases have addressed document-type classification and metadata completeness [4], not the characters of the abstract. Two relevant facts are documented. Several large publishers, including Elsevier and the American Chemical Society, deposit no abstracts to Crossref [5,6]. OpenAlex does not distribute plain-text abstracts and serves them only as an inverted index from which the text must be reconstructed [7]. For PubMed, the National Library of Medicine states that MEDLINE XML is UTF-8 in Unicode Normalization Form C and permits Greek letters and mathematical symbols [8]; normalization form C leaves dashes and curly quotation marks unchanged, and the page does not mention replacing them. To the author's knowledge, no study has compared the character-level fidelity of the major biomedical APIs against a common reference.

# 3. Methods

## 3.1 Overview of the audit method

The audit measures whether the characters of a reference text survive in the text that a service returns for the same record. Its five steps are stated here in general form.

1. Fix a list of code points, each assigned to a class, before any data are collected.
2. For every record, take a reference version of the text (here PMC JATS) and the text served by each service, matched by identifier.
3. Extract both with frozen rules and record, for each record and class, whether the class is present in the reference and in the served text.
4. Estimate two quantities that keep fidelity apart from availability, conditional preservation and end-to-end preserved availability, and classify each service × class cell by a preregistered interval rule.
5. For each cell classified as a loss, check the mechanism in the raw responses: the character at the aligned position and, for a service that returns an index, the raw index keys.

The rest of this section (§3.2–3.9) describes how these steps were applied in this study.

## 3.2 Preregistration

The study was registered on OSF Registries on 4 June 2026 at 04:16 UTC (https://osf.io/zjevr, Open-Ended Registration), before the sampling frame was downloaded (05:33 UTC) and before any analytic data were collected. An earlier pilot, with papers selected for the presence of target characters, informed the sample size and is described in the registration; pilot data are not part of this analysis. The registration froze, by SHA-256 hash, the list of 51 target code points with their class assignments (`p1_codepoints_v1.csv`) and the extraction code (`p1_extract_v1.py`, with a 36-assertion test suite); neither was modified afterwards. Every point that the registration left open, and every difference between registration and report, is listed in Supplementary Table S4. All random draws used seed 20260603, and all APIs were queried on a single date, 4 June 2026.

## 3.3 Sampling frame and sample

The sampling frame was the complete PMC Open Access file list (`oa_file_list.csv`, downloaded 4 June 2026, SHA-256 83bef966...), restricted to publication year 2024: 709,947 PMCIDs before filtering. Using the file list rather than the `esearch` endpoint avoids the undocumented pagination limit of that endpoint (`retstart` ≤ 9,998) and gives every record a known, equal probability of selection. Papers were included if the JATS `article-type` was `research-article` and the article language was English (JATS `xml:lang`, with the MEDLINE language field as fallback).

Papers were taken in the order of a single random permutation of the frame. As registered, a pre-screen of the first 500 included papers set the sample size: if the prevalence of Greek letters (the rarest primary class) were below 9%, N would be 5,000, otherwise 4,000. The observed prevalence was 12.6% (63/500), so N = 4,000. The analytic sample comprised the next 4,000 included papers in the same permutation, with no overlap with the pre-screen. No stratification or weighting was used.

The target population is English-language research articles in the 2024 PMC Open Access subset, not the biomedical literature as a whole. Because all four APIs are compared against the same JATS reference, between-API comparisons do not depend on how faithfully JATS itself reproduces the author's manuscript (§5.8).

### 3.4 Code points and character classes

Fifty-one code points were tracked in six classes, each defined a priori. Four classes are primary: (1) typographic punctuation: the hyphen U+2010, non-breaking hyphen U+2011, en dash U+2013, em dash U+2014, minus sign U+2212 and curly quotation marks U+2018, U+2019, U+201C, U+201D; (2) mathematical and scientific symbols: ±, ×, ≠, ≤, ≥, °, the micro sign U+00B5, Å and ‰; (3) Greek letters: twelve lower- and upper-case letters, including Greek mu U+03BC, which is distinct from the micro sign; and (5) special whitespace: the no-break space U+00A0, thin space U+2009, hair space U+200A and narrow no-break space U+202F. Two classes are exploratory: (4) Unicode subscript and superscript digits, contrasted with structural `<sup>/<sub>` markup (research question 4, RQ4), and (6) a miscellaneous class (bullet, ellipsis, ®, ©). The minus sign belongs to class 1 only, as registered.

### 3.5 Extraction

For JATS, the main `<abstract>` under `<article-meta>` was taken, excluding `<trans-abstract>` and graphical abstracts; sections of structured abstracts were concatenated, including their labels and titles. MathML content was not counted as text. HTML and XML entities were decoded before comparison, so that `—` counts as U+2014. PubMed `<AbstractText>` elements were concatenated in order. The Crossref abstract was stripped of markup without collapsing Unicode whitespace. The OpenAlex abstract was reconstructed from `abstract_inverted_index` by ordering words by position and joining them with a single U+0020 space; in addition, the presence of each code point in the raw index keys was recorded. The Semantic Scholar abstract was taken as plain text. An API abstract was eligible when it was non-empty after whitespace stripping; failures to map a PMCID to a PMID or DOI (one to three papers per source) and empty responses entered the coverage analysis, not the fidelity analysis.

### 3.6 Estimands, intervals and decision rule

For each API × class cell, conditional preservation (RQ1a) is the share of eligible papers (JATS-positive for the class and with a non-empty API abstract) in which the API abstract contains at least one code point of the class. End-to-end preserved availability (RQ1b) has all JATS-positive papers as its denominator, so that a missing abstract counts as not preserved. The metric is defined at the class level. The primary interval is the 95% Wilson interval; the Clopper–Pearson interval is a sensitivity analysis. A cell was classified as systematic loss when the upper bound was below 5%, faithful when the lower bound exceeded 90%, and partial otherwise; a verdict required at least 75 eligible papers. The systematic-loss label additionally required the mechanism audit (§3.8). As a sensitivity analysis only, one-sided binomial tests (H0: preservation ≥ 5%) with Benjamini–Hochberg correction were computed over the 16 primary cells.

### 3.7 Coverage and missing abstracts

Coverage (RQ2) is the share of papers for which an API returned a non-empty abstract, with Crossref coverage broken down by publisher. Because missing Crossref abstracts depend on the publisher, the Crossref end-to-end estimand is also bounded with worst-case Manski bounds, assuming 0% and 100% preservation in papers without an abstract.

### 3.8 Mechanism audit (RQ3)

For each systematic-loss cell, 50 discordant records (JATS-positive, character absent from the API abstract; seed 20260603) were drawn. For each record, the extraction code located the character in the JATS abstract, aligned the surrounding text with the API abstract and produced an evidence summary: for PubMed, the ASCII character found at the aligned position; for OpenAlex, whether the character occurred in the raw index keys. An auditor then assigned one mechanism from a fixed set: substitution, deletion, HTML/XML stripping, serialization loss, field-specific normalization, absence upstream in the API's source, or ambiguous. The auditor saw the character, the JATS snippet, the aligned API excerpt and the evidence summary, which named the replacing character or, for OpenAlex, the label "serialization". The audit therefore checked a programmatic classification against the text rather than classifying without it. Records were labeled only "A" or "B", with the key sealed until coding was complete; because each cell contained one API and one character class, blinding concerned the name of the API only. The characters at issue cannot be told apart by eye: a no-break space renders like an ordinary space. Asking a human to audit them seemed unpromising, so the registered auditor was a language model reading code points: Claude Haiku 4.5 (claude-haiku-4-5-20251001), run twice at temperature 0, which gives the registered intra-rater Cohen's kappa. A loss was retained when at least 90% of audited records received a mechanism other than "ambiguous". As an unregistered exploratory check, the same records were coded by four further models (GPT-4o, Llama-3.1-70B-Instruct, Mistral Large, DeepSeek-Chat; temperature 0). A fifth model (Gemini 2.0 Flash) was dropped by the rule fixed in the panel script before it ran, because more than half of its responses could not be parsed.

### 3.9 Verification

Before the preprint was posted, the verdicts of all 16 primary cells were recomputed with an independently written implementation and agreed in 16 of 16 cells. For a random subsample of 120 papers, the abstracts were re-extracted from the raw API responses with an independently written parser. Agreement on code-point presence in JATS was 4,079 of 4,080 checks; the one disagreement was a graphical abstract that the frozen extractor correctly excluded. There were no disagreements in preservation by any API. Analyses used Python with statsmodels.

## 4. Results

Of the 4,000 sampled papers, the JATS abstract contained at least one character of the primary classes as follows: typographic punctuation 2,114 (52.9%), special whitespace 1,086 (27.2%), mathematical and scientific symbols 639 (16.0%) and Greek letters 448 (11.2%). Every cell exceeded the minimum of 75 eligible papers. Table 1 and Figure 1 give conditional preservation for all 16 primary cells; Table 2 gives coverage. Clopper–Pearson intervals changed no verdict (Supplementary Table S2).

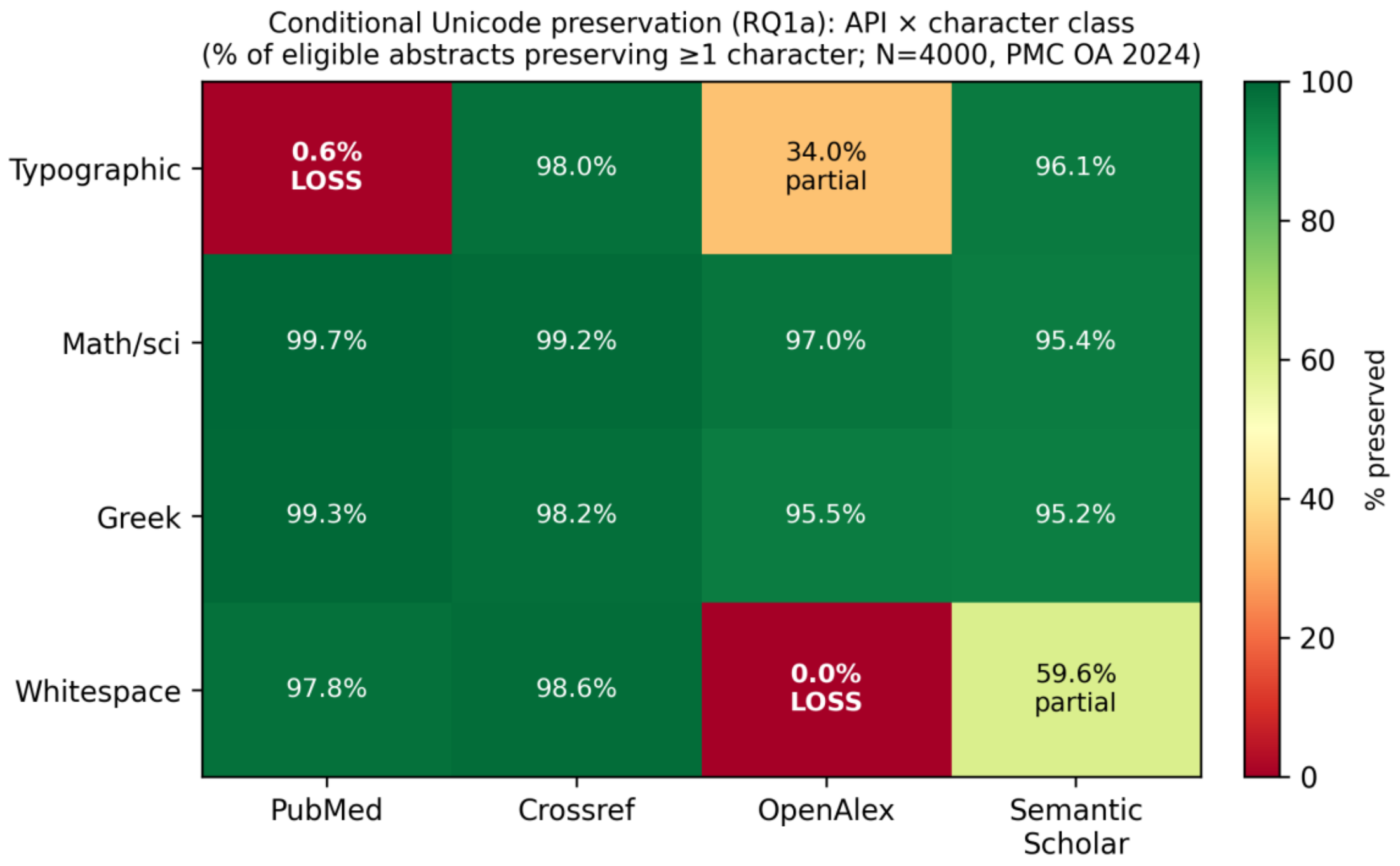

**Figure 1. Conditional preservation of Unicode characters (RQ1a), API × character class.** Share of eligible abstracts preserving at least one character of the class (N = 4,000, PMC Open Access 2024). Cells of systematic loss are labeled "LOSS": PubMed × typographic punctuation (0.6%) and OpenAlex × special whitespace (0.0%).

### 4.1 Two systematic losses

Two of the 16 cells met the systematic-loss criterion, and both had been predicted in the registration.

**PubMed × typographic punctuation.** PubMed preserved typographic punctuation in 0.6% of 2,113 eligible abstracts (12/2,113; 95% CI 0.3–1.0%; Benjamini–Hochberg-adjusted $p \approx 1.1 \times 10^{-30}$). En dashes, em dashes, Unicode hyphens, minus signs and curly quotation marks present in JATS were returned as their ASCII counterparts (U+002D, U+0027, U+0022).

**OpenAlex × special whitespace.** OpenAlex preserved special whitespace in 0.0% of 1,085 eligible abstracts (0/1,085; 95% CI 0.0–0.4%; adjusted $p \approx 5.4 \times 10^{-24}$). No special-whitespace code point occurred in any key of the raw inverted index. For every other code point, presence in the raw keys and presence in the reconstructed text agreed in all 2,389 paper–code-point pairs in which either occurred. The loss therefore arises in the index, not in reconstruction: an index of words and positions has no place for the characters between words.

## 4.2 Mathematical symbols and Greek letters are preserved by all four APIs

All eight cells for mathematical symbols and Greek letters were faithful (lower 95% bound > 90%). Preservation ranged from 95.4% (Semantic Scholar) to 99.7% (PubMed) for mathematical symbols and from 95.2% (Semantic Scholar) to 99.3% (PubMed) for Greek letters. Characters that carry meaning in biomedical text, such as β in "β-blocker", ± in a confidence interval and μ in "μg", therefore pass through all four APIs.

## 4.3 Partial cells

Two cells were partial: OpenAlex × typographic punctuation, 34.0% (715/2,101; 95% CI 32.0–36.1%), and Semantic Scholar × special whitespace, 59.6% (642/1,078; 95% CI 56.6–62.4%). Because characters present in the raw OpenAlex keys were always retained on reconstruction (§4.1), the missing typographic characters were already absent from the index. The registered prediction that Crossref and Semantic Scholar would preserve characters at high rates held for Crossref in all four classes and for Semantic Scholar in three; whitespace in Semantic Scholar was partial.

**Table 1. Conditional preservation (RQ1a), 4 APIs × 4 primary classes.** Preserved / eligible (%), [95% Wilson CI], verdict.

| Class | PubMed | Crossref | OpenAlex | Semantic Scholar |
|---|---|---|---|---|
| Typographic | 12/2,113 (0.6%) [0.3–1.0] **loss** | 1,648/1,682 (98.0%) [97.2–98.5] faithful | 715/2,101 (34.0%) [32.0–36.1] partial | 2,006/2,088 (96.1%) [95.2–96.8] faithful |
| Math/science | 636/638 (99.7%) [98.9–99.9] faithful | 479/483 (99.2%) [97.9–99.7] faithful | 617/636 (97.0%) [95.4–98.1] faithful | 604/633 (95.4%) [93.5–96.8] faithful |
| Greek | 445/448 (99.3%) [98.0–99.8] faithful | 320/326 (98.2%) [96.0–99.2] faithful | 426/446 (95.5%) [93.2–97.1] faithful | 419/440 (95.2%) [92.8–96.9] faithful |
| Whitespace | 1,061/1,085 (97.8%) [96.7–98.5] faithful | 766/777 (98.6%) [97.5–99.2] faithful | 0/1,085 (0.0%) [0.0–0.4] **loss** | 642/1,078 (59.6%) [56.6–62.4] partial |

## 4.4 Coverage and the Crossref deposit gap

PubMed returned an abstract for 99.35% of papers (95% CI 99.05–99.56%), OpenAlex for 99.43% (99.14–99.62%) and Semantic Scholar for 98.33% (97.88–98.68%). Crossref returned an abstract for 75.4% (3,017/4,000; 95% CI 74.1–76.7%), so one paper in four had none. The gap depended on the publisher (Table 3): MDPI, Frontiers, Oxford University Press, PLOS and SAGE deposited abstracts for all sampled papers, Springer Nature for 75.0%, and Elsevier (0/360) and the American Chemical Society (0/105) for none, consistent with earlier reports [5,6]. The publisher breakdown is descriptive; the full table is in Supplementary Table S1.

**Table 2. Coverage per API (RQ2).**

| API | Non-empty / N | Coverage | 95% CI |
|---|---|---|---|
| PubMed | 3,974 / 4,000 | 99.35% | 99.05–99.56 |
| OpenAlex | 3,977 / 4,000 | 99.43% | 99.14–99.62 |
| Semantic Scholar | 3,933 / 4,000 | 98.33% | 97.88–98.68 |
| Crossref | 3,017 / 4,000 | 75.43% | 74.07–76.73 |

**Table 3. Crossref abstract coverage by publisher group (selected).**

| Publisher group | n | Coverage |
|---|---|---|
| MDPI / Frontiers / OUP / PLOS / SAGE | 1,416 | 100% |
| Wiley | 218 | 99.1% |
| Wolters Kluwer (Ovid) | 117 | 96.6% |
| Springer Nature | 1,042 | 75.0% |
| Other | 659 | 61.8% |
| Elsevier | 360 | 0% |
| American Chemical Society | 105 | 0% |

## 4.5 Conditional and end-to-end estimands

For Crossref, the two estimands diverge. When an abstract is present, Crossref is faithful in all four classes (98.0–99.2%). Counting papers without a Crossref abstract as not preserved, end-to-end preservation falls to 78.0% for typographic punctuation, 75.0% for mathematical symbols, 71.4% for Greek letters and 70.5% for special whitespace. Manski bounds, which make no assumption about the missing abstracts, place the Crossref end-to-end estimand for typographic punctuation in [78.0%, 98.4%] and for whitespace in [70.5%, 99.0%]. For the other three APIs, end-to-end and conditional estimates differed by at most 1.7 percentage points (Supplementary Table S2).

## 4.6 Mechanism audit

The programmatic evidence located both losses. In 45 of the 50 audited PubMed records, the aligned position held an ASCII character: the hyphen-minus U+002D (33 records), the apostrophe U+0027 (10) or the quotation mark U+0022 (2); in the remaining 5, the code found no aligned counterpart. In all 50 audited OpenAlex records, the special space was absent from the raw index keys. The registered audit confirmed both losses as deterministic (Table 4). For PubMed × typographic punctuation, the auditor assigned substitution to 50 of 50 records. For OpenAlex × special whitespace, it assigned serialization loss to 42 and substitution to 8, so all 50 records had a non-ambiguous mechanism. Both labels describe the same observation, a special space returned as U+0020; given the inverted-index format (§4.1), serialization loss is the accurate one. Under a stricter reading that requires a single mechanism in at least 90% of records, the registered auditor's 84% would not meet the threshold, whereas each model of the exploratory panel assigned serialization loss to 49 or 50 of 50 records. Cohen's kappa between the two passes was 1.000. Both losses were retained.

In the exploratory panel, each of the four models classified 100% of records in both cells as deterministic and agreed with the registered auditor's dominant mechanism; agreement with the registered auditor's record-level codes was 91–92%, and Fleiss' kappa across the panel was 0.977. The registration required that, until a mechanism was confirmed, the losses be described only as differences from JATS; with the mechanisms confirmed, they are described below as normalization by the API.

**Table 4. Mechanism audit of the two systematic-loss cells (n = 50 records each; auditor blinded to the API name).**

| Cell | Mechanism (registered auditor) | Non-ambiguous | Intra-rater kappa |
|---|---|---|---|
| PubMed × typographic | substitution 50/50 | 100% | 1.000 |
| OpenAlex × whitespace | serialization loss 42/50, substitution 8/50 | 100% | 1.000 |

Exploratory four-model panel, both cells combined (100 records): Fleiss' kappa 0.977.

## 4.7 Single characters

Class-level verdicts hide differences between characters. Table 5 gives conditional preservation for the most frequent characters (descriptive).

**Table 5. Conditional preservation of selected single characters (%), ordered by JATS prevalence.**

| Character | Papers with character in JATS | PubMed | Crossref | OpenAlex | Semantic Scholar |
|---|---|---|---|---|---|
| – en dash U+2013 | 1,082 (27%) | 0.6 | 97.6 | 36.9 | 95.2 |
| ’ right single quotation mark U+2019 | 862 (22%) | 0.2 | 97.7 | 24.5 | 95.5 |
| thin space U+2009 | 661 (17%) | 98.6 | 98.6 | 0.0 | 93.8 |
| no-break space U+00A0 | 590 (15%) | 95.9 | 96.9 | 0.0 | 8.4 |
| − minus sign U+2212 | 252 (6%) | 0.8 | 98.0 | 62.2 | 92.8 |
| ‐ hyphen U+2010 | 204 (5%) | 0.0 | 96.5 | 34.3 | 87.6 |
| — em dash U+2014 | 104 (3%) | 1.0 | 96.4 | 26.9 | 92.3 |

The characters PubMed loses most often are the en dash (27% of papers) and the apostrophe (22%); replacing a curly apostrophe with a straight one changes nothing for a reader. The partial whitespace result for Semantic Scholar combines two behaviors: the thin space was preserved (93.8%) and the no-break space was not (8.4%). From a human reader’s point of view, the losses fall into three groups:

- **No change in meaning:** apostrophes and quotation marks, the minus sign and special spaces replaced by ASCII equivalents. These make up most of the losses.
- **Loss for programmatic matching:** the en dash and Unicode hyphen in ranges and compounds; “300–500 mg” retrieved from PubMed becomes “300-500 mg”, which a pattern or dictionary keyed on the en dash no longer matches.
- **Loss for measurement:** the em dash carries little meaning but is studied as a marker of LLM-assisted writing (§5.5).

## 4.8 Exploratory analyses

Structural <sup>/<sub> markup occurred in 736 of 4,000 JATS abstracts (18.4%) and Unicode superscript or subscript digits in 8 (0.2%), as the registration predicted for RQ4.

The miscellaneous class occurred in 195 abstracts, mostly as bullets (156); only 54 of these papers had a Crossref abstract. Excluding records with a preprint DOI changed no verdict. Details are in Supplementary Table S3.

# 5. Discussion

## 5.1 Principal findings

The same paper does not reach every tool in the same form. Depending on the API, its abstract has typographic punctuation replaced by ASCII (PubMed), special spaces replaced by ordinary ones (OpenAlex), its characters preserved (Crossref, when an abstract exists) or no abstract at all (Crossref, for some publishers). To a reader, "300–500" and "300-500" are the same; even here, it takes a second look to see that they differ. Reviewers and authors therefore do not notice such differences, while every machine that reads the literature at scale receives them.

## 5.2 Consequences for biomedical text mining and corpora

Three consequences follow for pipelines that build corpora or extract information from abstracts. First, corpora built from Crossref abstracts are missing whole publishers (Table 3), so they are not a random subset of the literature. Second, the same paper retrieved from two APIs yields different byte sequences, hence different tokens and, without prior normalization, different embeddings and hashes; deduplication, semantic search and benchmark reproducibility are affected when sources are mixed. The divergence in representation is certain; its effect on task performance was not measured. Third, literal matching (dictionaries, gazetteers, regular expressions and text-hash deduplication) fails under substitution: an extractor that recognizes dose ranges by the en dash finds almost none of them in PubMed abstracts (Table 5).

## 5.3 Harm depends on the consumer

The grouping in §4.7 grades the losses by what they mean to a human reader, although the question concerns machine reading. For a machine, the harm can be graded by how certain it is. The change in representation is certain: every substitution changes the bytes, the tokens and, in pipelines without Unicode normalization, the embedding or hash. The effect on deterministic operations is also certain: anything that matches, deduplicates or counts the literal string diverges. The effect on model behavior is open: whether a model interprets the variants differently was not tested. For a human reader, most of what PubMed changes is cosmetic; the consequential losses are limited to the en dash and Unicode hyphen for matching and the em dash for measurement.

## 5.4 Causes and remedies of the two losses

The two losses arise at different points. PubMed's substitution was observed in the `<AbstractText>` field, the only PubMed field examined; it is deterministic and is reversible in principle because the publisher's JATS keeps the original characters; the remedy lies with the National Library of Medicine, which could preserve the characters or document the substitution. OpenAlex's loss of special whitespace follows from its format: an inverted index has no representation for the characters between words, so no reconstruction can recover them, and the remedy is a different format or at least an explicit caveat. For users, this means that typographic punctuation should be taken from Crossref or Semantic Scholar, mathematical symbols and Greek letters can be taken from any of the four APIs, and special whitespace should not be taken from an inverted-index source. Whatever the source, the API and retrieval date should be reported, like a software version.

## 5.5 Consequences for studies of AI-assisted writing

Character-level markers of AI-assisted writing depend on the API that serves the text. The em dash, whose rise has been reported in abstracts retrieved from OpenAlex [2], survived in 1.0% of eligible PubMed abstracts, 26.9% in OpenAlex and 96.4% in Crossref; its frequency in PubMed abstracts therefore reflects the National Library of Medicine's processing as much as the authors' writing. Judged by the em dash alone, PubMed abstracts would be almost free of language models. Lexical markers such as those of Kobak et al. [1] consist of whole words, which the substitutions observed here do not touch. Any character-level marker should be validated against the serving API before it is read as a sign of authorship. The en dash, ten times more frequent than the em dash in these abstracts, should be counted separately from it. In the author's companion study, measuring the em dash required leaving PubMed and analyzing full-text XML of medRxiv preprints [9].

## 5.6 What this study does not claim

The study does not claim that normalization impairs plagiarism detection, which was not studied. It does not claim lower recall in search engines for human users, whose handling of punctuation was not studied; the harm to literal matching concerns programmatic operations. It does not claim that language models perform worse on normalized text; the findings concern the infrastructure, and any effect on model output would need a separate experiment.

## 5.7 The audit method: generalizability and range of applicability

Beyond its findings, the study contributes a method for auditing the character-level fidelity of any service that returns text which also exists in a reference version. The method (§3.1) rests on four parts, all fixed before data collection: a frozen list of code points with class assignments; a common reference text for every record (here PMC JATS); two estimands that keep fidelity apart from availability; and a mechanism check in the raw responses (the character at the aligned position, the raw index keys). To the author's knowledge, current practice has no such check: studies take the text as their chosen source returns it [1–3], and comparisons of databases measure coverage or metadata completeness [4,5], so a loss of characters is invisible to both. The data of this study show what each approach can see. Measured by coverage alone, as in earlier database comparisons [4,5], PubMed and OpenAlex look nearly complete: they returned an abstract for 99.35% and 99.43% of papers (Table 2). The audit shows that the same two sources lose typographic punctuation (PubMed, preserved in 0.6% of eligible abstracts) and special whitespace (OpenAlex, 0.0%) (Table 1).

The method applies wherever a reference text can be matched to the served text through an identifier: other bibliographic APIs, other fields such as titles, later retrieval dates, or other sets of characters, for example diacritics in authors' names. Adding an API requires a fetcher and an extraction rule for that API; the instrument, the estimands and the decision rule stay unchanged. The range is limited by the reference. Records without an open full text, which is most of the literature outside PMC Open Access, cannot be audited this way, and the reference may itself be normalized (§5.8). The class-level "any survival" metric is coarse by design; for a single character of interest, the per-character level (§4.7) is the right one.

## 5.8 Limitations

First, JATS is the reference, not the author's manuscript: publishers may normalize characters during typesetting, so JATS is an upper bound for what any API could return, and between-API comparisons are unaffected. Second, the population is the 2024 PMC Open Access subset of English-language research articles; publishers outside PMC Open Access may deposit differently. Third, API behavior was observed on a single date; the registered check of stability across publication years was not performed (Supplementary Table S4). Fourth, missing Crossref abstracts depend on the publisher, so they were bounded rather than imputed. Fifth, the mechanism audit adds little independent evidence: the auditor saw an evidence summary that named the replacing character or the serialization label, and an intra-rater kappa of 1.000 for a model run twice at temperature 0 reflects the determinism of the model, not human-grade reliability. The mechanisms rest mainly on the programmatic evidence, namely the ASCII character at the aligned position in PubMed and the absence of special spaces from the raw OpenAlex index keys. Sixth, the metric is class-level "any survival"; per-character results are descriptive.

## 5.9 Future work

Still open: whether PubMed's substitution is stable over time and across ingestion pipelines; how often authors use ASCII in the manuscript before any pipeline touches it; and whether the demonstrated divergence in representation changes the output of search systems or language models trained or prompted with these abstracts. The same audit could be pointed at clinical data pipelines, where text passes from source systems through interfaces into warehouses and registries and where a lost character can change meaning: a dose in µg that arrives as "ug" or "g" is no longer the same dose. In the four bibliographic APIs studied here, the micro sign survived in 94–100% of the 64 abstracts that contained it; whether it survives on its way through a hospital's data pipeline was not studied.

## 6. Conclusion

Bibliographic APIs do not return the same characters for the same abstract. PubMed replaces typographic punctuation with ASCII, OpenAlex cannot carry special whitespace, Crossref lacks abstracts from several large publishers, and all four preserve Greek letters and mathematical symbols. Character-level analyses of biomedical abstracts measure the serving pipeline as well as the literature unless the source is validated and reported.

The findings apply to anyone who builds a corpus or extracts information from abstracts retrieved through these APIs. In practice: choose the source according to the characters the analysis needs, normalize all sources in the same way when they are mixed, and report the API and the retrieval date. The audit itself, a fixed list of code points compared against a common reference, can be rerun at a later date with the published code and extended to other APIs by adding a fetcher.

---

## Declarations

**CRediT authorship contribution statement.** Przemysław Czuma: Conceptualization, Methodology, Software, Validation, Formal analysis, Investigation, Data curation, Writing – original draft, Writing – review and editing, Visualization, Project administration.

**Declaration of competing interest.** The author declares no competing financial interests. Non-financial interests: the author is the founder of the Polish Association for Artificial Intelligence in Medicine (inteligencja.org.pl), which advocates the use of artificial intelligence, including large language models, to improve human health. The author is also developing a free, non-commercial LLM-based tool that summarizes medical literature; that work is the source of the author's familiarity with programmatic retrieval of biomedical articles, which informed the technical approach taken here. The tool was not used to obtain or process the data in this study and played no role in its design, analysis or interpretation.

**Funding.** This research did not receive any specific grant from funding agencies in the public, commercial, or not-for-profit sectors.

**Ethics.** The study used only published bibliographic records and involved no human participants; ethical approval was not required.

**Data availability.** The per-paper analytic data, codebook and checksums are archived on Zenodo [10]. The analysis code is available at https://github.com/P-Czum/pmc-unicode-audit (MIT license). The preregistration and its frozen attachments are at https://osf.io/zjevr. The PMC Open Access file list used as the sampling frame is public NCBI data and is not redistributed (snapshot of 4 June 2026, SHA-256 83bef966a0490916c5864a5d2be883eac03b9afe1473144356f28cce049690d4).

**Use of language models as a research instrument.** Language models served as the source-blinded mechanism auditor and as an exploratory panel (§3.8); models, versions and settings are reported there.

## Declaration of generative AI and AI-assisted technologies in the manuscript preparation process

During the preparation of this work the author used Claude (Anthropic) to assist with writing analysis code, checking statistical computations and editing the language of the manuscript. After using this tool, the author reviewed and edited the content as needed and takes full responsibility for the content of the published article.

---

# Supplementary material. Invisible to humans, visible to machines: a preregistered audit of Unicode fidelity across four biomedical bibliographic APIs

## Table S1. Crossref abstract coverage by publisher group (full)

Publisher group from the Crossref `publisher` field; groups below 2% of the sample are combined as “Other”. Descriptive.

| Publisher group | n | Coverage |
|---|---|---|
| Springer Science and Business Media LLC | 1,042 | 75.0% |
| MDPI AG | 748 | 100.0% |
| Other | 659 | 61.8% |
| Elsevier BV | 360 | 0.0% |
| Frontiers Media SA | 287 | 100.0% |
| Wiley | 218 | 99.1% |
| Oxford University Press (OUP) | 146 | 100.0% |
| Public Library of Science (PLoS) | 142 | 100.0% |
| Ovid Technologies (Wolters Kluwer Health) | 117 | 96.6% |
| American Chemical Society (ACS) | 105 | 0.0% |
| SAGE Publications | 93 | 100.0% |
| openRxiv | 83 | 100.0% |

## Table S2. End-to-end preserved availability (RQ1b) and Clopper–Pearson intervals for conditional preservation (RQ1a)

RQ1b: preserved / all JATS-positive papers (%), [95% Wilson CI]. RQ1a Clopper–Pearson: 95% exact interval for the conditional estimate in Table 1 of the main text. No verdict changes under Clopper–Pearson intervals.

| Class | API | RQ1b | RQ1a, Clopper–Pearson 95% CI |
|---|---|---|---|
| Typographic | PubMed | 12/2,114 (0.6%) [0.3–1.0] | 0.3–1.0 |
| Typographic | Crossref | 1,648/2,114 (78.0%) | 97.2–98.6 |

| Class | API | RQ1b | RQ1a, Clopper–Pearson 95% CI |
|---|---|---|---|
| | | [76.1–79.7] | |
| Typographic | OpenAlex | 715/2,114 (33.8%) [31.8–35.9] | 32.0–36.1 |
| Typographic | Semantic Scholar | 2,006/2,114 (94.9%) [93.9–95.8] | 95.1–96.9 |
| Math/science | PubMed | 636/639 (99.5%) [98.6–99.8] | 98.9–100.0 |
| Math/science | Crossref | 479/639 (75.0%) [71.5–78.2] | 97.9–99.8 |
| Math/science | OpenAlex | 617/639 (96.6%) [94.8–97.7] | 95.4–98.2 |
| Math/science | Semantic Scholar | 604/639 (94.5%) [92.5–96.0] | 93.5–96.9 |
| Greek | PubMed | 445/448 (99.3%) [98.0–99.8] | 98.1–99.9 |
| Greek | Crossref | 320/448 (71.4%) [67.1–75.4] | 96.0–99.3 |
| Greek | OpenAlex | 426/448 (95.1%) [92.7–96.7] | 93.2–97.2 |
| Greek | Semantic Scholar | 419/448 (93.5%) [90.9–95.5] | 92.8–97.0 |
| Whitespace | PubMed | 1,061/1,086 (97.7%) [96.6–98.4] | 96.7–98.6 |
| Whitespace | Crossref | 766/1,086 (70.5%) [67.8–73.2] | 97.5–99.3 |
| Whitespace | OpenAlex | 0/1,086 (0.0%) [0.0–0.4] | 0.0–0.3 |
| Whitespace | Semantic Scholar | 642/1,086 (59.1%) [56.2–62.0] | 56.6–62.5 |

Manski bounds for the Crossref end-to-end estimand: typographic [78.0%, 98.4%]; math/science [75.0%, 99.4%]; Greek [71.4%, 98.7%]; whitespace [70.5%, 99.0%].

## Table S3. Exploratory analyses

**RQ4, Unicode versus structural sub- and superscripts (registered).** Structural <sup>/<sub> markup: 736 of 4,000 JATS abstracts (18.4%). Unicode superscript or subscript digits (class 4): 8 of 4,000 (0.2%; superscript two in 7, superscript three in 1); 4 of these also contained structural markup. Class 4 cells had at most 8 eligible papers and received no verdict.

**Class 6, miscellaneous (registered, descriptive).** JATS-positive papers: 195 (bullet 156, registered sign 37, ellipsis 1, copyright sign 1). Conditional preservation: PubMed 61/195 (31.3%); Crossref 49/54 (90.7%); OpenAlex 82/193 (42.5%); Semantic Scholar 82/193 (42.5%). Only 54 of the 195 papers had a Crossref abstract.

**OpenAlex raw index keys versus reconstructed text (registered).** For every paper–code-point pair in which a target code point occurred in either the raw `abstract_inverted_index` keys or the reconstructed text (2,389 pairs), it occurred in both. No special-whitespace code point (class 5) occurred in any raw key.

**Preprint records (not registered).** Excluding the 85 papers with a preprint DOI (prefix 10.1101) changed none of the 16 primary verdicts; the largest change in a conditional estimate was 1.2 percentage points (Crossref × Greek, 98.2% → 99.4%).

## Table S4. Registration compared with report

Registration: OSF osf.io/zjevr, 4 June 2026, 04:16 UTC. “As registered” means performed as specified.

| Registered item | Report | Where |
|---|---|---|
| RQ1a, RQ1b, RQ2 | As registered | §4.1–4.5; Tables 1–3, S2 |
| RQ3 mechanism audit and retention rule | As registered; see clarifications below | §3.8, §4.6 |
| RQ4 sub/superscripts | As registered | §4.8; Table S3 |
| Pre-screen rule (Greek < 9% → N = 5,000) | As registered; observed 12.6% (63/500), N = 4,000 | §3.3 |
| Pre-screen “before the main draw” | Implemented as the first 500 included papers of one permutation, with the analytic sample taken from the following papers (no overlap); decided before the frame was downloaded | §3.3 |
| False discovery rate (FDR) sensitivity analysis | Clarification: registration did not state the null hypothesis; one-sided binomial test, H0: preservation ≥ 5%, Benjamini–Hochberg over 16 cells | §3.6 |
| Mechanism auditor | Clarification: registration | §3.8 |

| Registered item | Report | Where |
|---|---|---|
| | says “auditor” without specifying a human; a language model was used (Claude Haiku 4.5, two passes at temperature 0) | |
| Auditor input | Not specified in the registration: each record included an evidence summary generated by the extraction code, naming the replacing character (PubMed) or the label “serialization” (OpenAlex) | §3.8, §5.8 |
| Intra-rater coding “after a time gap” | Two independent passes of a stateless model; no time gap | §3.8 |
| Retention threshold “deterministic mechanism in ≥ 90%” | Clarification: read as ≥ 90% non-ambiguous records. Under a single-mechanism reading, OpenAlex × whitespace reaches 84% with the registered auditor | §4.6 |
| Multi-model panel | Not registered; exploratory. One of five models dropped by the pre-set parsing rule | §3.8, §4.6 |
| Crossref publisher breakdown (exploratory) | As registered | §4.4; Table S1 |
| Manski bounds | As registered | §4.5; Table S2 |
| Missing-at-random-within-publisher (MAR) sensitivity (Crossref) | Not performed | n/a |
| Domain subgroup (exploratory) | Not performed; the domain field was not populated at collection | n/a |
| Miscellaneous class (exploratory) | As registered | Table S3 |
| OpenAlex raw-token analysis (exploratory) | As registered | §4.1; Table S3 |
| Temporal stability 2021–2024 (exploratory) | Not performed | §5.8 |

| Registered item | Report | Where |
| --- | --- | --- |
| Per-character preservation | Not registered; descriptive | §4.7 |
| Preprint-record sensitivity | Not registered | Table S3 |
| Release of registration, code and data together with the preprint | The preprint was posted on 8 June 2026; the code repository and the registration are public; the Zenodo data record was published on 22 November 2026 | Data availability |